\documentclass[journal=jacsat,manuscript=article]{achemso}

\usepackage[version=3]{mhchem} 
\usepackage{xcolor}

\author{Jaros\l{}aw Paw\l{}owski}
\affiliation{Institute of Theoretical Physics, Wroc\l{}aw University of Science and Technology, Wroc\l{}aw, Poland}
\email{jaroslaw.pawlowski@pwr.edu.pl}
\author{Pankaj Kumar}
\affiliation[National University of Singapore]
{Institute for Functional Intelligent Materials, National University of Singapore, Singapore}
\alsoaffiliation[National University of Singapore]
{Integrative Sciences and Engineering Programme, National University of Singapore, Singapore}
\author{Kenji Watanabe}
\affiliation[NIMS]
{Research Center for Functional Materials, National Institute for Materials Science, Tsukuba, 305-0044, Japan}
\author{Takashi Taniguchi}
\affiliation[NIMS]
{Research Center for Functional Materials, National Institute for Materials Science, Tsukuba, 305-0044, Japan}
\author{Konstantin S Novoselov}
\affiliation[National University of Singapore]
{Institute for Functional Intelligent Materials, National University of Singapore, Singapore}
\alsoaffiliation[National University of Singapore]
{Integrative Sciences and Engineering Programme, National University of Singapore, Singapore}
\alsoaffiliation[National University of Singapore]
{Department of Materials Science and Engineering, National University of Singapore, Singapore}
\author{Hugh O.~H.~Churchill}
\affiliation[University of Arkansas]
{Department of Physics, University of Arkansas, Fayetteville, Arkansas}
\alsoaffiliation[MonArk]
{MonArk NSF Quantum Foundry, University of Arkansas, Fayetteville, Arkansas, 72701, USA}
\author{Dharmraj Kotekar-Patil}
\affiliation[University of Arkansas]
{Department of Physics, University of Arkansas, Fayetteville, Arkansas}
\email{dk030@uark.edu}
\alsoaffiliation[MonArk]
{MonArk NSF Quantum Foundry, University of Arkansas, Fayetteville, Arkansas, 72701, USA}
\alsoaffiliation[A*STAR]
{Institute of Materials Research and Engineering (IMRE), Agency
for Science, Technology and Research (A*STAR), Singapore 138634}

\title[An \textsf{achemso} demo]
  {Single Electron Quantum Dot in Two-Dimensional Transition Metal Dichalcogenides}

\abbreviations{IR,NMR,UV}
\keywords{American Chemical Society, \LaTeX}

\begin{document}

\begin{abstract}
Spin-valley properties in two-dimensional (2D) semiconducting transition metal dichalcogenides (TMDC) has attracted significant interest due to the possible applications in quantum computing. 
Spin-valley properties can be exploited in TMDC quantum dot (QD) with well-resolved energy levels.
This requires smaller QDs, especially in material systems with heavy carrier effective mass e.g. TMDCs and silicon.  
Device architectures employed for TMDC QDs so far have difficulty achieving smaller QDs.
Therefore, an alternative approach in the device architecture is needed. 
Here, we propose a multilayer device architecture to achieve a gate-defined QD in TMDC with a relatively large energy splitting on the QD.
We provide a range of device dimensions and dielectric thicknesses and its correlation with the QD energy splitting.
The device architecture is modeled realistically. 
Moreover, we show that all the device parameters used in modeling are experimentally achievable.
These studies lay the foundation for future work toward spin-valley qubits in TMDCs.  
The successful implementation of these device architectures will drive the technological development of 2D materials-based quantum technologies.
\end{abstract}

\section{Introduction}

Two-dimensional (2D) semiconducting transition metal dichalcogenides (TMDCs) are atomically thin semiconducting materials. 
The ultrathin body of the TMDCs provides strong confinement and electrostatic control along one of its crystal axis (out of the plane axis). 
This makes them an ideal platform for confining individual charges and spins.  
TMDCs are particularly interesting for spin-valley qubits due to their unique combination of electronic and optical properties, which make them well-suited for manipulating and controlling both the spin and valley degrees of freedom.
Single-layer TMDCs have broken inversion symmetry resulting in non-identical valleys at the K and K’ points of the Brillouin zone. 
This results in charge carriers acquiring an additional quantum number known as their valley index or valley pseudo-spin at the K and K valleys. 
This individual valley addressability combined with spins can be utilized in new qubit architectures. 
Heavy transition metal atoms in TMDCs introduce strong spin-orbit coupling (SOC). 
SOC with large opposite sign spin-splitting at the K and K’ valleys near the band edges leads to the coupling of spin and valley. 
Consequently, the scattering of charge carriers between the valleys demands a large momentum transfer between K and K’ in addition to the simultaneous spin flip. 
This is expected to be a slow process, resulting in possibly long coherence times\cite{xu_spin_2014}.
However, the development of this field in TMDC is in its infancy and further research is required to compare the spin coherence times with advanced material systems such as GaAs and silicon. 
The first step towards measuring spin coherence times in a TMDC QD device is a demonstration of QD with well-resolved excited states, which can serve as two two-level systems of the qubit. 
Recent theoretical work proposes the possibility of using spin and valley states in electrostatically gated QDs and nanoribbons in TMDCs as qubits \cite{Pawlowski2018,Pawlowski2019,Pawlowski2021,Menaf2021}.

Electrostatically defined QDs in TMDCs has been demonstrated by multiple group \cite{pisoni_gate-tunable_2018,wang_electrical_2018,zhang_electrotunable_2017,boddison-chouinard_gate-controlled_2021, davari_gate-defined_2020}. 
The reported experiments employ a set of planar gates to define QD in TMDC.
QDs achieved in most experiments operate in the so-called "metallic" regime where the energy separation on the QD ($\Delta$) is smaller than the thermal energy ($k_B T$), ie $\Delta$ $<$ $k_B T$.
Experiments demonstrating smaller QDs with relatively large energy splitting has mostly replied on sample inhomogeneity and defects to study the fundamental properties in TMDCs \cite{papadopoulos_tunneling_2020,devidas_spectroscopy_2021,kumar_excited_nodate}.
However, in materials such as TMDCs where the effective mass (m$_{eff}=0.3$-$0.7$~m$_0$) is much higher m$_0=$~free electron mass), this imposes a stringent requirement on the size of the QD to observe the level spacings on the QD.
In comparison with silicon in its development phase, it suffered from similar problems in achieving a tigher confinement using the planar gate architecture.
The solution in silicon was achieved by shrinking the host material (silicon) into nanostructure such as etched nanowires \cite{hofheinz_simple_2006}, lithographically defined QD \cite{leobandung_observation_1995} or by spliting the confinement gates into different fabrication layers (often referred to as "overlapping" gates) \cite{angus_gate-defined_2007}.
Here, we adopt a similar strategy by splitting the gates in two different device layer.
The device architecture proposed here provides a co-relation between the device dimensions and the QD energy spectrum, providing a clear pathway towards designing a few electron QD necessary for the implementation of spin-valley qubits.

In this work, we propose a multilayer device architecture to create an electrostatically defined QD in TMDCs.
To achieve a tighter confinement, the device architecture relies on splitting the set of gates in different fabrication layers.
We built a realistic model for the proposed device and performed numerical simulations.
Based on the simulations, we propose device dimensions (gate dimensions and dielectric thickness) to enhance the $\Delta$ and show that the device dimensions used for the model are experimentally achievable in the labs. 


\section{Device Structure}
The artistic image of the proposed device architecture is shown in Figure~1(a). 
Figure~1(b) shows the cross-section of the device displaying different device layers.
The device consists of several components, such as source and drain contact leads, encapsulated TMDC between boron nitride, and a set of gates.
The device uses a heavily doped silicon substrate covered with an 285 nm thick silicon dioxide (SiO$_2$). 
Doped silicon along with SiO$_2$ acts as a global backgate that can be used to globally tune the carrier density in the TMDC. 
To create a strong confinement in TMDC, two sets of gates are used. 
These gates are split into two different layers.
First, the bottom split gates with a gap between them (SG) is placed on SiO$_2$ (red region in Figure~1(c)).
Then, the TMDC which is encapsulated between two boron nitride flakes (hBN/TMDC/hBN) is placed on top of the prepatterned bottom split gates.
The three fine gates (FG) are placed on top of the encapsulated stack (hBN/TDC/hBN) overlapping with the bottom SG (shown in Figure~1(c)).
Splitting SG and FG into two different layers allows the same TMDC channel region to be electrostatically gated, creating a compact device architecture and creating a tighter confinement (Figure~1(c)).
Finally, the source and drain electrodes are defined on top of the TMDC forming Ohmic contacts.
To avoid suppression of charge injection at cryogenic temperatures due to the Schottky barrier, source and drain contacts need to be Ohmic at cryogenic temperatures.

\begin{figure}
\includegraphics[width=16cm]{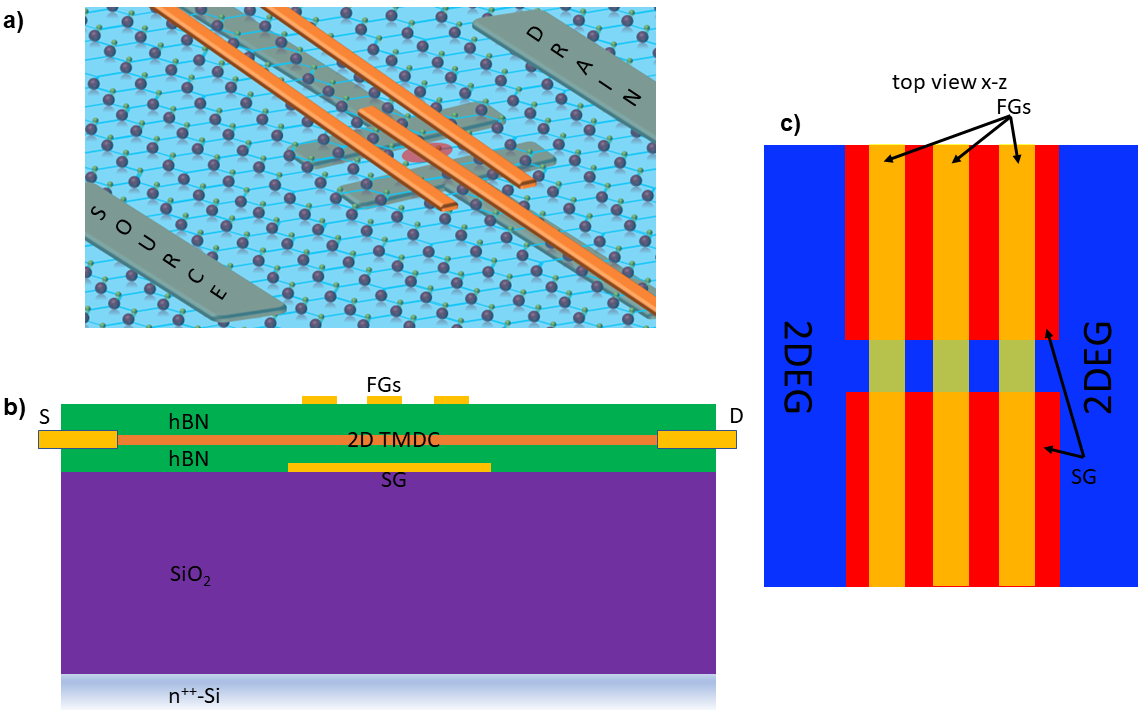}
\caption{Schematic of the device architecture: a) Artistic representation of the device architecture proposed to access the few-electron regime in TMDC. The schematic exhibits an encapsulated 2D TMDC between two layers of boron nitride. Two layers of gates overlapping in two different fabrication layers separated by an encapsulated stack of hBN/2D TMDC/hBN. Bottom split gates confines the two-dimensional electron gas into a one-dimensional electron gas. Top fine gates further confine the one-dimensional electron gas into zero-dimension electron gas (QD). Source and drain electrodes form Ohmic contact, enabling to measure current through the device. b) Schematic of the proposed device cross-section shown in panel a). c) Top view of the proposed device architecture with accumulated two-dimensional electron gas. Split gate and the top fine gates can be used to tune the quantum dot and barriers in the device to achieve a QD size comparable to the fine gate dimensions.}
\centering
\end{figure}

Two sets of gates, SG and FG play an important role in depleting the two-dimensional electron gas (2DEG) to a zero-dimension electron gas (0D) - QD.
This can be implemented in two steps.
First, the SG gate depletes 2DEG by applying the appropriate voltage that leaves charge carriers only between the SG gate, forming a one-dimensional electron gas (1DEG) (Figure~2(a)).
Then, by applying appropriate voltages to the three FGs, the 1DEG can be tuned into a 0D QD (Figure~2(b)).
The outer two FGs define the barrier region whereas the central FG controls the QD potential.
By tuning the three FGs one can control the barriers and the QD potential profile independently.
The electrochemical potential profile from a shallow QD to a deep QD can be formed along the xy axis, as shown in Figure~2(c) and 2(d).


\section{Theoretical model with self-consistent solutions}

\begin{figure}[bt]
\includegraphics[width=16cm]{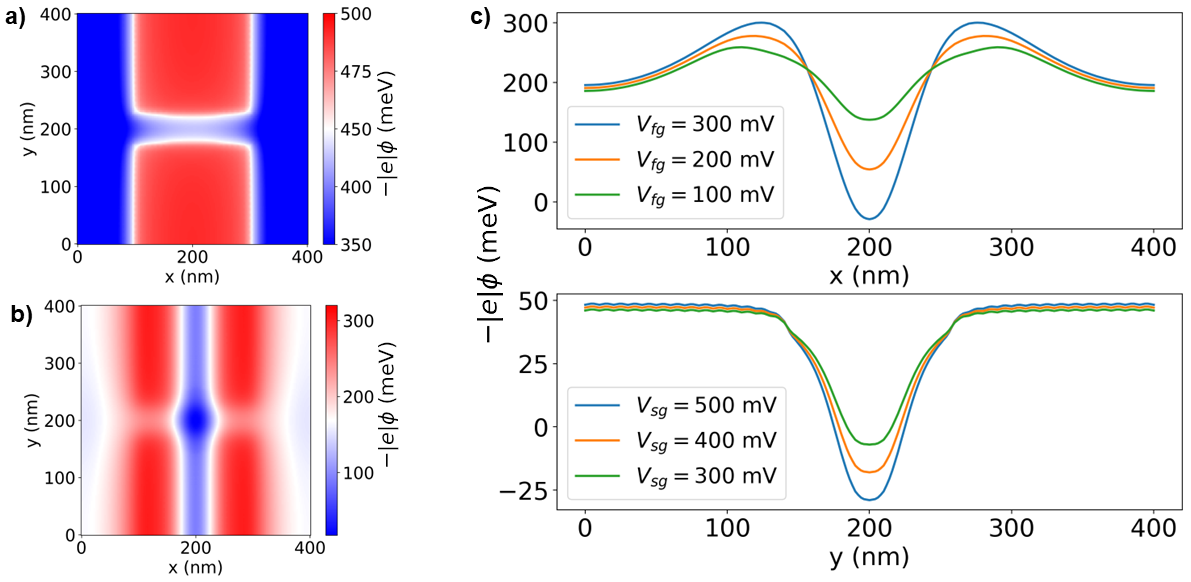}
\centering
\caption{\label{fig:pot} Potential landscape seen at the TMDC layer. 
(a) Potential profile at the TMDC level before applying potential to FGs (only SGs active with $V_\mathrm{sg}=500$~mV) -- negative voltage on split gates confine 2DEG into a 1DEG. 
(b) Potential profile at the TMDC level after applying voltages to FGs ($V_\mathrm{fg}=\pm300$~mV) -- elongated QD emerges at the center of the device. 
(c) Potential line cuts along $x$-axis for variable $V_\mathrm{fg}$ values (top, $V_\mathrm{sg}=500$~mV) showing high confinement tunability, and along $y$-axis with variable $V_\mathrm{sg}$ (bottom, $V_\mathrm{fg}=\pm300$~mV). 
Results for the assumed SG gap of $50$~nm; and tunnel barriers width of $20$~nm.}
\end{figure}

To model single-electrons confined in the in-plane monolayer QD an effective-mass model have been used. 
The model takes into account states in the vicinity of the conduction band minimum (at \textit{K} and \textit{K'} valley), within spin $s$ and valley $\mathcal{K}$ degenerate subspace\cite{Burkard2014,Liu2014}.
The continuum 2D Hamiltonian in this basis is of a form\cite{Pawlowski2018}
\begin{equation}\label{eq:ham}
H^\mathrm{2D}(x,y)=
 \begin{pmatrix}
  \frac{p^2}{2m_{\uparrow K}}+\varphi+\Delta_\mathrm{SO} & 0 & \frac{\lambda_R}{\hbar} p_{-} & 0 \\
  0 &  \frac{p^2}{2m_{\uparrow K'}}+\varphi-\Delta_\mathrm{SO} & 0 & \frac{\lambda_R}{\hbar} p_{-} \\
  \frac{\lambda_R}{\hbar} p_{+} & 0 & \frac{p^2}{2m_{\downarrow K}}+\varphi-\Delta_\mathrm{SO} & 0 \\
  0 & \frac{\lambda_R}{\hbar} p_{+} & 0 & \frac{p^2}{2m_{\downarrow K'}}+\varphi+\Delta_\mathrm{SO}
 \end{pmatrix},
\end{equation}
with squared momentum operator $p^2=-\hbar^2(\partial_x^2+\partial_y^2)$, $p_{\pm}=p_x\pm\imath p_y$, and momentum components $p_i\equiv-\imath\hbar\partial_i$. 
The 2D potential energy is defined using 3D confinement potential: $\varphi(x,y)=-|e|\phi(x,y,z_{ML})$, where $z_{ML}$ is the monolayer position on $z$-axis in the nanodevice.

The Hamiltonian (\ref{eq:ham}) includes four subbands with spin-valley states
$\{|s,\mathcal{K}\rangle
=|s\rangle\otimes|\mathcal{K}\rangle\}=\{
|\!\uparrow\!K \rangle,
|\!\uparrow\!K'\rangle,
|\!\downarrow\!K\rangle,
|\!\downarrow\!K'\rangle\}$, forming the basis. The spin-valley levels are further split by $2\Delta_\mathrm{SO}=3$~meV,\cite{Kosmider2013} the intrinsic spin-orbit interaction.
The effective masses $m_{s,\mathcal{K}}$ for the set of four subbands fulfils the relation: $m_{s,-\mathcal{K}}=m_{-s,\mathcal{K}}$, and the assumed band masses are $m_{\uparrow K}=m_{\downarrow K'}=0.49$~$m_e$, and $m_{\uparrow K'}=m_{\downarrow K}=0.44$~$m_e$.\cite{Burkard2014}
This model also allows to include the Rashba type spin-orbit interaction, induced by an external perpendicular electric field $E_z$ and with spin orbit coupling $\lambda_R=\gamma_R|e|E_z$, where $\gamma_R=3.3\cdot10^{-4}$~nm$^2$ and the field is expressed in V~nm$^{-1}$.\cite{Burkard2014} 

Voltages applied to the gates (relative to the substrate) are used to create the QD confinement potential in the monolayer area. 
The confined electron itself carries charge density (which we account for via the mean field approach).
To calculate the realistic electrostatic potential $\phi(\mathbf{r})$ we solve the generalized Poisson equation~\cite{Pawlowski2016}:
\begin{equation}\label{eq:poiss}
\begin{split}
    \boldsymbol{\nabla}\!\cdot\left(\varepsilon_0\varepsilon(\mathbf{r})\boldsymbol{\nabla}\Phi(\mathbf{r})\right)=-\rho_e(\mathbf{r}),\\
    \phi(\mathbf{r})=\Phi(\mathbf{r})-\phi_e(\mathbf{r}),
\end{split}
\end{equation}
taking into account all of these components: voltages $V_\mathrm{sg}$ and $V_\mathrm{fg}$ applied to the control gates and to the highly doped substrate (kept at the referential potential $V_\mathrm{bg}=0$), space-dependent permittivity $\varepsilon(\mathbf{r})$ of different materials in the device (we assume $\varepsilon_\mathrm{MoS_2}=6.4$, $\varepsilon_\mathrm{SiO_2}=3.9$, and $\varepsilon_\mathrm{hBN}=3.3$)~\cite{Laturia2018}, and electron charge density $\rho_e(\mathbf{r})$ itself.
At the lateral and top sides of the computational box we apply Neumann boundary conditions with zeroing normal component of the electric field. 
To remove electron self-interaction we subtract electron potential itself $\phi_e(\mathbf{r})$ from the total potential $\Phi(\mathbf{r})$. Electron potential $\phi_e(\mathbf{r})$ is calculated using the standard Poisson equation: $\boldsymbol{\nabla}^2\phi_e(\mathbf{r})=-\rho_e(\mathbf{r})/(\varepsilon_0\varepsilon(\mathbf{r}))$ without any voltage conditions on the gates or the substrate.
Knowing 3D confinement potential also enables us to calculate electric field 
$E_z=-\partial_z\varphi(\mathbf{r})$.
Further details of the used method can be found in~[\!\!\citenum{Pawlowski2018}]. 

To calculate the electron eigenstates we solve the Schr\"odinger equation for the Hamiltonian~(\ref{eq:ham}) with electrostatic potential $\phi(\mathbf{r})$,
\begin{equation}\label{eq:schr}
    H^\mathrm{2D}[\phi(x,y,z_{ML})]\Psi(x,y)=E\Psi(x,y).
\end{equation}
To solve the Schr\"odinger equation (\ref{eq:schr}) we had to discretize the Hamiltonian (\ref{eq:ham}) on 2D square lattice, translating partial derivatives into finite differences. On the other hand, to calculate the electrostatic potential one must solve the Poisson equation~(\ref{eq:poiss}), which in turn requires knowledge of the electron charge density in a given QD state $\rho_e(\mathbf{r})=-|e||\Psi(x,y)|^2\eta(z-z_{ML})$.
with Gaussian charge distribution $\eta$ centered at $z=z_{ML}$.
This means that both of the equations need to be solved self-consistently.
Within the model we include four spin-valley degenerate subbands, therefore, full wave function (envelope) in this representation takes the form 
\begin{equation}\label{eq:spinor}
\Psi(x,y) = [
\psi_{\uparrow K},
\psi_{\uparrow K'},
\psi_{\downarrow K},
\psi_{\downarrow K'}
]^T.
\end{equation}

\begin{figure}[t]
\includegraphics[width=16cm]{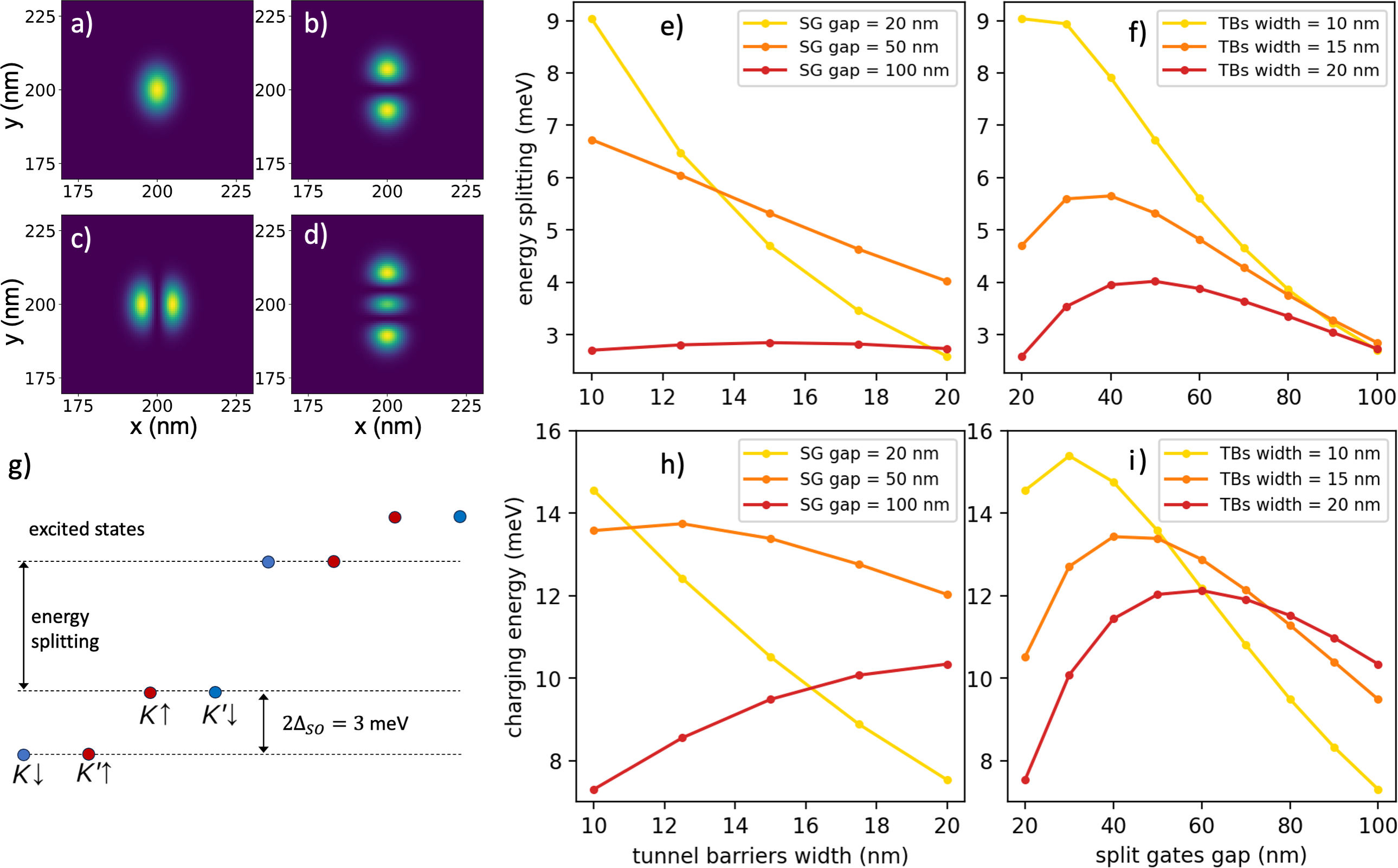}
\centering
\caption{\label{fig:dens} (a-d) Electron density for the ground and first few excited states  with different orbital configuration. Energy splitting between ground state and first excited state (denoted in (g)) with different orbital configuration for variable tunnel barriers width (e), and variable split gates gap (f). Spin-valley qubit subspace is separated from the excited states by the spitting energy as explained in (g). Additionally, QD charging energies were esimated (h,i) for the same device parameter variations (TBs width and SG gap)  as in (e,f). Assumed voltages are: $V_L=300$, $V_C=-300$, $V_R=300$ (fine gates), $V_\mathrm{sg}=500$, $V_\mathrm{back}=-2000$~(mV).}
\end{figure}

Presented in Figure~\ref{fig:dens} electron densities (a-d) for QD-confinement from Figure~\ref{fig:pot} shows ground state density $|\Psi(x,y)|^2$ and its first few excited states with different spatial part calculated via the Poisson-Schr\"odinger system defined in Eqs.~\ref{eq:poiss} and \ref{eq:schr}.
In next steps, we studied density and energy levels with orbital excitation because at this stage of the device designing we are mainly interested in the confinement shape optimization. Energy difference (splitting) that separates spin-valley qubit subspace from the excited energy levels, as presented in Figure~\ref{fig:dens}(g), is the objective of device geometry optimization and the goal is to have this energy gap high.
Therefore we identify the energy splitting between ground state and first excited state with different orbital configuration as a parameter characterising confinement potential. The stronger -- deeper the QD confinement is, the higher energy splitting we obtain.
Before we discuss the results of the numerical model, we introduce the nomenclature used in reference to the device.
$V_\mathrm{fg}$ and $V_\mathrm{sg}$ refers to the voltages applied on the SG and FG gates.
SG gaps refers to the spacing between the two SG gates.
Tunnel barrier (TB) widths are the thickness of the hBN used in the device that acts as a gate dielectric and energy splitting is the level spacing on the QD.

Using the developed numerical model, we can optimize the nanodevice structure in terms of the strongest confinement.
Strong confinement is associated with larger energy splitting.
In Figure~\ref{fig:dens}(e) presents the dependence of energy splitting on the width of the tunnel barriers for different SG gaps. 
It is generally perceived that the narrow gate dimensions provide stronger confinement, however consideration must be given to the gate dielectric thickness (referred to as tunnel barrier width here).
For example, in Figure~\ref{fig:dens}(e), SG gap = 20~nm provides largest energy splitting when TB = 10~nm but provides lowest energy splitting when TB increases to 20~nm.
The diverging electric fields with increasing TB widths affects the confinement as the voltages applied on SG gates needs to be adjusted to maintain a 1D channel between them while depleting the 2DEG underneath them.
In that regard, a SG gap of 50~nm provides stronger confinement with thicker TB compared to smaller SG gap.
It is clear that thinner tunnel barrier yields sharper and deeper QD confinements as long as the SG gap is not too large (the optimum value is $< 50$~nm).
Thinner tunnel barriers yield sharper potentials which is based on the observation that at a distance, all potentials generated by applying boundary conditions (voltages) on gates get softened.
The simulations suggest using as thin a tunnel barrier as possible (but one that ensures no breakdown leakage) when building the nanodevice.

On the other hand, Figure~\ref{fig:dens}(f) shows the dependence of energy splitting on SG gap width for different tunnel barriers. We observe that there is optimal SG gap value (depending on TBs width), e.g. for TBs width of 20~nm the optimal SG gap should be about 50~nm -- and this is the configuration used in Figure~\ref{fig:pot}.
The observed drop in confinement strength (smaller energy spitting) for smaller SG gaps is due to the fact that a smaller SG gap causes the substrate to be more strongly screened by the split gates. In the limit where we have no SG gap, we will not get any confinement along the y-axis.
Hence, in addition to the the gate dimensions it is necessary to choose the TB widths appropriately.
Experimental evidence of thinner TB widths on energy splitting is demonstrated in Ref.\cite{davari_gate-defined_2020}, which is one of the few reports demonstrating excited states in gate defined quantum dot.

There are two important energy scales present in the system when considering an electron, confined in the QD, as a potential qubit host. First one, the single-particle energy splitting, is determined by the confinement potential strength and describes how the qubit subspace is separated from the rest of the system -- see Figure~\ref{fig:dens}(g): the higher the energy splitting is, the lower chance that the electron will be excited out of the qubit subspace during the adiabatic manipulation. The second one, here equals spin-orbit splitting $2\Delta_{SO}$ (but can be fine-tuned by applying external magnetic field) -- see Figure~\ref{fig:dens}(g) and the Hamiltonian~(Eq.~\ref{eq:ham}), describes internal energy scale of the qubit (in case of a qubit defined separately on spin or valley degrees), or separates lower spin-valley (Kramers) pair from their SOI split-off partners (in case of spin-valley qubit). In general, to get a good candidate for a qubit, its internal energy spacing should be much smaller than the distance to higher orbital states. And in our case, the confinement provides promising energy spiting from the higher subbands.

One of the essential parameters of a QD is the charging energy~\cite{Kouwenhoven_2001,david_effective_2018} – the energy cost of adding another electron to the system. We have estimated capacitance $C$ of the formed QD (capacitance mainly between QD and the nearby split and fine gates) using the Poisson solver by calculating the charge induced on the QD for a given gate-defined electrostatic environment. Afterwards, we can calculate the charging energy as~\cite{Kouwenhoven_2001} $E_c=e^2/C$, which is presented in Figure~\ref{fig:dens}(h,i). Similar method was applied for estimating the charging energy $E_c$ in graphene QD~\cite{freitag} or InAs nanowires~\cite{kretinin}. The obtained $E_c$ energies are typically 2-3 times larger than the separation between single-particle states (energy splitting) as presented in Figure~\ref{fig:dens}(e,f). 

\begin{figure}
\includegraphics[width=12cm]{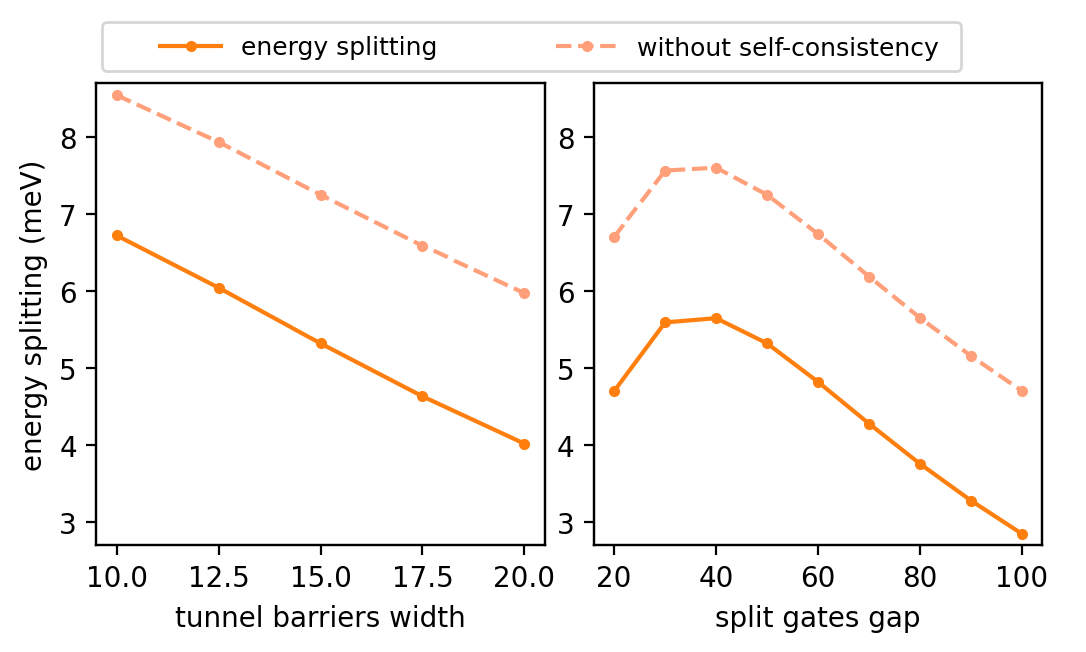}
\centering
\caption{\label{fig:self} Self-consistency in calculations is important to get correct energy spitting values. Presented calculations for energy splitting vs. TBs width for SG gap~$=50$~nm (left), and energy splitting vs. SG gap for TBs width $=15$~nm (right).}
\end{figure}

One should also note that if we turn off self-consistency by putting $\rho_e(\mathbf{r})\equiv0$ in the Poisson equation (\ref{eq:poiss}) we arrive at very different (typically much overestimated) energy splitting values, as presented in Figure~\ref{fig:self}.
Therefore, this constraint in calculations, although complicates and elongates simulation time, it is important to get correct description of the QD confinement and estimations of the splitting.

In this way, one can optimize any other geometric dimensions of the device, or test other arrangements of the control gates layout or their $z$-position. In the Supplementary Materials we present and discuss results for some other arrangements that were tested and gave slightly worse confinements. We analyzed other stackings of the fine gates, split gates, and monolayer; optimize voltages for such arrangements, and calculate energy splitting for the electron confined therein. We also propose a configuration that enables to obtain the double QD structure.

\section{Device dimensions and electrical properties}

The theoretical model provides a range of device dimensions (SG gap, FG gate dimensions and tunnel barrier widths (gate dielectric thickness)) and their relationship to the energy splitting on the QD.
To successfully implement the device architecture, we need three device components:

1. Ohmic contact at cryogenic temperatures

2. Gate dielectric thickness (tunnel barrier widths)

3. Gate dimensions (SG and FG)

While the theoretical model does not discuss the role of source and drain contacts explicitly, it assumes the Ohmic nature of contacts at cryogenic temperatures.
One of the key challenges TMDC-based quantum devices face is poor Ohmic contact, especially at cryogenic temperature. 
A few reports have successfully demonstrated low contact resistance devices at cryogenic temperatures but often require a large backgate voltage\cite{cui_multi-terminal_2015,cui_low-temperature_2017}. 
A large backgate voltage applied fills the TMDC channel with charge carriers, which makes it difficult to deplete the channel with local gates to reach few charge carriers on the QD.
The metal deposition process using evaporator has been shown to induce defects in TMDCs that cause a Fermi-level pinning\cite{liu_approaching_2018}. 
Over the years, different strategies have been developed to form Ohmic contact and the van der Waals type of contact is considered the most suitable without causing any damage to 2D materials.
vdWs contacts can be achieved in different ways e.g. stacking TMDCs with source-drain electrodes (metal or graphene) to form a heterostructure or using low melting temperature metals as source-drain contacts using standard evaporation technique \cite{liu_approaching_2018,wang_van_2019,shen_ultralow_2021,pisoni_interactions_2018}.

Here we show that these approaches can be successfully implemented for TMDC devices to achieve Ohmic contact at cryogenic temperatures.
Figure~5(a) shows an artistic view of heterostructure device using single layer MoS$_2$.
VdW heterostructure was prepared using a single layer MoS$_2$ contacted by a few-layer graphite (FLG).
This SL MoS$_2$/FLG heterostructure was encapsulated between two hexagonal boron nitride (hBN) flakes (Figure~5(a)).
The whole stack was assembled inside Ar/N$_2$ filled glovebox with oxygen concentration $<1$~ppm. 
FLG was contacted using a Cr/Au electrodes through one-dimensional edge contacts\cite{wang_one-dimensional_2013}.
A relatively thin hBn flakes ($\sim10$~nm) were used to encapsulate the heterostructure.
This is the thinnest TB widths used in the model.
The 10~nm hBN flake was initially chosen by optical contrast and later verified using a cross-section TEM image.
A narrow top gate was placed on the stack as schematically shown in Figure~5(a).
In this device, we demonstrate the first two conditions mentioned above (Ohmic contact and thin tunnel barrier) to implement the proposed device architecture.

\begin{figure}
\includegraphics[width=16cm]{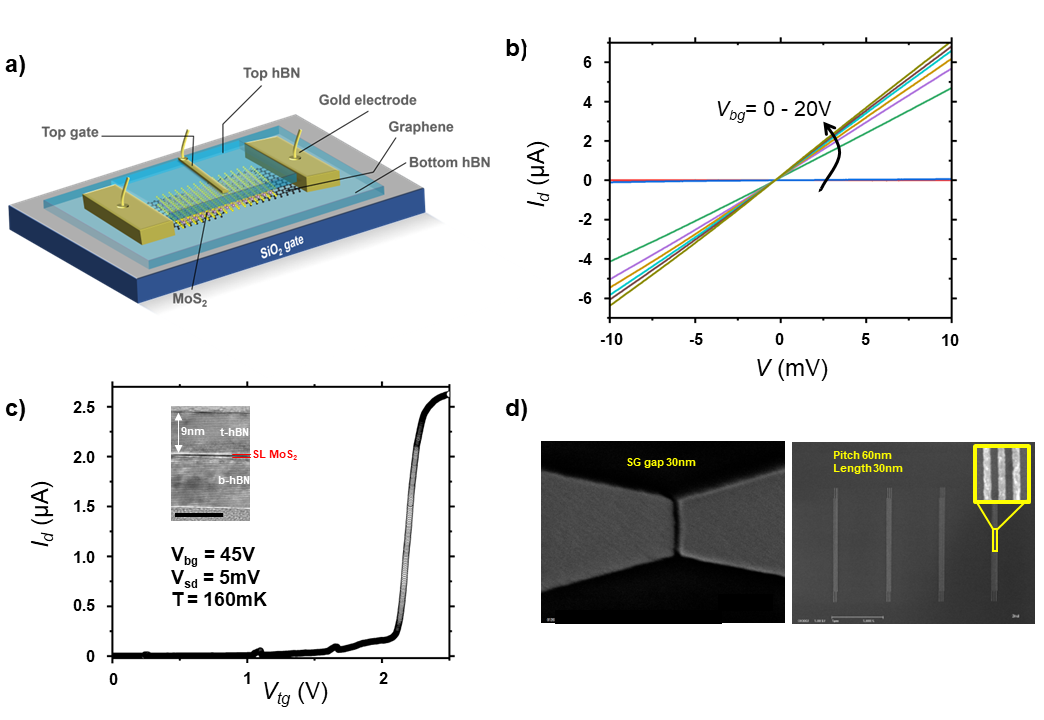}
\caption{(a) Schematic of a van der Waals heterostructure using single layer MoS$_2$ encapsulated between hBN flakes. (b) $I_d$--$V$ for varying backgate voltage $V_\mathrm{bg}=0$--$20$~V at $T=160$~mK in a single layer MoS$_2$ contacted by few-layer graphene. The device exhibits linear $I_d$--$V$ traces, a characteristic of Ohmic contact. Two-point resistance of $\sim1.4$~k$\Omega$ at deep cryogenic temperature ensuring a transparent MoS$_2$ and FLG interface. (c) $I_d$--$V_\mathrm{tg}$ trace for a device shown in panel (a). Inset shows a TEM cross-section image of the heterostructure exhibiting thickness of different device layers. (d) \textit{Left:} Scanning electron microscope image SG gates with a gap of 30~nm between them. \textit{Right:} Scanning electron microscope image of set of FG gates with a pitch of 60~nm and gate length of 30~nm.}
\end{figure}

Figure~5(b) shows the output transfer characteristic of the device at $T=160$~mK.
The different color $I$--$V$ traces correspond to the varying back-gate voltage $V_\mathrm{bg}=0$--$20$~V. 
Linear $I$--$V$ traces at such a low temperature ensure the Ohmic nature of our contacts.
Relatively small $V_\mathrm{bg}$ range and fast switching between OFF-state to ON-state is a consequence of high mobility in our device.
The two-point resistance extracted from Figure~5(b) exhibits a device resistance of 1.4~k$\Omega$, which includes twice the contact resistance and the channel resistance.
This sets the upper limit on contact resistance in our SL MoS$_2$ device to $\sim 700$~$\Omega$.
Such a low contact resistance at deep cryogenic temperature further validates the suitability of vdWs contact for quantum devices in TMDCs.
Similarly, vdWs contact between a low melting metal (bismuth) and single layer MoS$_2$ showing Ohmic nature is shown in supplementary Figure~S7.

To confirm that the TBs width used in the modeling can provide good electrostatic control over the channel, we measure $I_d$ vs $V_\mathrm{tg}$ at $T=160$~mK as shown in Figure~5(c).
For the lower $V_\mathrm{tg}$ range ($<1.7$~V), the $I_d$ exhibits a few Coulomb oscillations originating from the sample inhomogeneity below the top gate.
With a further increase in $V_\mathrm{tg}$, the device rapidly switches from an insulating state to a metallic state for a higher $V_\mathrm{tg}$ range ($>$ 2V).
This shows that the hBN thickness of $\sim10$~nm (thinest TB width used in the device modeling) provides excellent electrostatic control.

Finally, the gate dimensions (SG and FG) used in the device modeling are in the few tens of nanometer range (optimal $<50$~nm).
Figure~5(d)~(left) shows the scanning electron micrograph (SEM) of an SG gate fabricated with a gap of 30~nm using electron beam lithography and metalized with Cr/Au.
Since gate SG lies at the bottom of the device stack, the roughness (or smoothness) of these gates plays a critical role in device quality.
One solution is to employ a patterned trilayer graphene using a resist-free local anodic oxidation of graphite technique using atomic force microscope which can produce atomically flat surfaces \cite{cohen_nanoscale_2023}. 
Alternatively, thin layers of platinum has shown to produce smooth metal films that can serve as flat SG gates in the architecture \cite{davari_gate-defined_2020}.
Similarly, Figure~5(d)~(right) shows the SEM picture of the series of FGs with a gate length of 30~nm and pitch of 60~nm fabricated using EBL.
The device dimensions used in modeling are within the experimental reach using fabrication tools available in the research labs.


\section{Conclusion}

In summary, we propose a multi-layer device architecture to achieve a gate-defined quantum dot in TMDCs. 
Two sets of gates are used to create confinement that are placed in two different device layers, which provides a tighter confinement.
Using an effective-mass model, we perform device modelling to provide a co-relation between the device dimensions and energy splitting on the quantum dot.
This co-relation allows to choose from a range of different device dimensions to maximize the energy level splitting on the quantum dot.
Lastly, we show that the device dimensions used in the device modelling are experimentally achievable using standard fabrication processes in the research labs.
Future work will focus on implementing the full device architecture towards demonstration of gate-defined quantum dot in TMDCs.

\begin{acknowledgement}
This research was supported in part by PL-Grid Infrastructure and the Wroclaw Center for Networking and Supercomputing (WCSS). JP acknowledges support from National Science Centre, Poland, under grant no. 2021/43/D/ST3/01989.
This research is supported in part by the Ministry of Education, Singapore, under its Research Centre of Excellence award to the Institute for Functional Intelligent Materials (I-FIM, project No. EDUNC-33-18-279-V12). K. S. N. is grateful to the Royal Society (UK, grant number RSRP\textbackslash{}R\textbackslash{}190000) for support. K. W. and T. T. acknowledge the support from JSPS KAKENHI (Grant numbers 19H05790, 20H00354 and 21H05233).
DKP is grateful to the Agency for Science, Technology, and Research (\textit{A$^*$STAR}) for support under its \textit{A$^*$STAR} Career Development Fund (Grant number C210112039) and the MonArk NSF Quantum Foundry supported by the National Science Foundation Q-AMASE-i program under NSF award No. DMR-1906383.  HC acknowledges support from AFOSR award No.~FA9550-19-S-0003. 

\end{acknowledgement}




\bibliography{bibliography}

\providecommand{\latin}[1]{#1}
\makeatletter
\providecommand{\doi}
  {\begingroup\let\do\@makeother\dospecials
  \catcode`\{=1 \catcode`\}=2 \doi@aux}
\providecommand{\doi@aux}[1]{\endgroup\texttt{#1}}
\makeatother
\providecommand*\mcitethebibliography{\thebibliography}
\csname @ifundefined\endcsname{endmcitethebibliography}
  {\let\endmcitethebibliography\endthebibliography}{}
\begin{mcitethebibliography}{34}
\providecommand*\natexlab[1]{#1}
\providecommand*\mciteSetBstSublistMode[1]{}
\providecommand*\mciteSetBstMaxWidthForm[2]{}
\providecommand*\mciteBstWouldAddEndPuncttrue
  {\def\EndOfBibitem{\unskip.}}
\providecommand*\mciteBstWouldAddEndPunctfalse
  {\let\EndOfBibitem\relax}
\providecommand*\mciteSetBstMidEndSepPunct[3]{}
\providecommand*\mciteSetBstSublistLabelBeginEnd[3]{}
\providecommand*\EndOfBibitem{}
\mciteSetBstSublistMode{f}
\mciteSetBstMaxWidthForm{subitem}{(\alph{mcitesubitemcount})}
\mciteSetBstSublistLabelBeginEnd
  {\mcitemaxwidthsubitemform\space}
  {\relax}
  {\relax}

\bibitem[Xu \latin{et~al.}(2014)Xu, Yao, Xiao, and Heinz]{xu_spin_2014}
Xu,~X.; Yao,~W.; Xiao,~D.; Heinz,~T.~F. Spin and pseudospins in layered
  transition metal dichalcogenides. \emph{Nature Physics} \textbf{2014},
  \emph{10}, 343--350\relax
\mciteBstWouldAddEndPuncttrue
\mciteSetBstMidEndSepPunct{\mcitedefaultmidpunct}
{\mcitedefaultendpunct}{\mcitedefaultseppunct}\relax
\EndOfBibitem
\bibitem[Paw\l{}owski \latin{et~al.}(2018)Paw\l{}owski, \ifmmode~\dot{Z}\else
  \.{Z}\fi{}ebrowski, and Bednarek]{Pawlowski2018}
Paw\l{}owski,~J.; \ifmmode~\dot{Z}\else \.{Z}\fi{}ebrowski,~D.; Bednarek,~S.
  Valley qubit in a gated ${\mathbf{MoS}}_{2}$ monolayer quantum dot.
  \emph{Phys. Rev. B} \textbf{2018}, \emph{97}, 155412\relax
\mciteBstWouldAddEndPuncttrue
\mciteSetBstMidEndSepPunct{\mcitedefaultmidpunct}
{\mcitedefaultendpunct}{\mcitedefaultseppunct}\relax
\EndOfBibitem
\bibitem[Paw{\l}owski(2019)]{Pawlowski2019}
Paw{\l}owski,~J. Spin-valley system in a gated
  \emph{New Journal of Physics} \textbf{2019}, \emph{21}, 123029\relax
\mciteBstWouldAddEndPuncttrue
\mciteSetBstMidEndSepPunct{\mcitedefaultmidpunct}
{\mcitedefaultendpunct}{\mcitedefaultseppunct}\relax
\EndOfBibitem
\bibitem[Paw\l{}owski \latin{et~al.}(2021)Paw\l{}owski, Bieniek, and
  Wo\ifmmode~\acute{z}\else \'{z}\fi{}niak]{Pawlowski2021}
Paw\l{}owski,~J.; Bieniek,~M.; Wo\ifmmode~\acute{z}\else \'{z}\fi{}niak,~T.
  Valley Two-Qubit System in a ${\mathrm{Mo}\mathrm{S}}_{2}$-Monolayer Gated
  Double Quantum dot. \emph{Phys. Rev. Applied} \textbf{2021}, \emph{15},
  054025\relax
\mciteBstWouldAddEndPuncttrue
\mciteSetBstMidEndSepPunct{\mcitedefaultmidpunct}
{\mcitedefaultendpunct}{\mcitedefaultseppunct}\relax
\EndOfBibitem
\bibitem[Alt\ifmmode \imath \else \i \fi{}nta\ifmmode~\mbox{\c{s}}\else
  \c{s}\fi{} \latin{et~al.}(2021)Alt\ifmmode \imath \else \i
  \fi{}nta\ifmmode~\mbox{\c{s}}\else \c{s}\fi{}, Bieniek, Dusko,
  Korkusi\ifmmode~\acute{n}\else \'{n}\fi{}ski, Paw\l{}owski, and
  Hawrylak]{Menaf2021}
Alt\ifmmode \imath \else \i \fi{}nta\ifmmode~\mbox{\c{s}}\else \c{s}\fi{},~A.;
  Bieniek,~M.; Dusko,~A.; Korkusi\ifmmode~\acute{n}\else \'{n}\fi{}ski,~M.;
  Paw\l{}owski,~J.; Hawrylak,~P. Spin-valley qubits in gated quantum dots in a
  single layer of transition metal dichalcogenides. \emph{Phys. Rev. B}
  \textbf{2021}, \emph{104}, 195412\relax
\mciteBstWouldAddEndPuncttrue
\mciteSetBstMidEndSepPunct{\mcitedefaultmidpunct}
{\mcitedefaultendpunct}{\mcitedefaultseppunct}\relax
\EndOfBibitem
\bibitem[Pisoni \latin{et~al.}(2018)Pisoni, Lei, Back, Eich, Overweg, Lee,
  Watanabe, Taniguchi, Ihn, and Ensslin]{pisoni_gate-tunable_2018}
Pisoni,~R.; Lei,~Z.; Back,~P.; Eich,~M.; Overweg,~H.; Lee,~Y.; Watanabe,~K.;
  Taniguchi,~T.; Ihn,~T.; Ensslin,~K. Gate-tunable quantum dot in a high
  quality single layer {MoS2} van der {Waals} heterostructure. \emph{Applied
  Physics Letters} \textbf{2018}, \emph{112}, 123101\relax
\mciteBstWouldAddEndPuncttrue
\mciteSetBstMidEndSepPunct{\mcitedefaultmidpunct}
{\mcitedefaultendpunct}{\mcitedefaultseppunct}\relax
\EndOfBibitem
\bibitem[Wang \latin{et~al.}(2018)Wang, De~Greve, Jauregui, Sushko, High, Zhou,
  Scuri, Taniguchi, Watanabe, Lukin, Park, and Kim]{wang_electrical_2018}
Wang,~K.; De~Greve,~K.; Jauregui,~L.~A.; Sushko,~A.; High,~A.; Zhou,~Y.;
  Scuri,~G.; Taniguchi,~T.; Watanabe,~K.; Lukin,~M.~D.; Park,~H.; Kim,~P.
  Electrical control of charged carriers and excitons in atomically thin
  materials. \emph{Nature Nanotechnology} \textbf{2018}, \emph{13},
  128--132\relax
\mciteBstWouldAddEndPuncttrue
\mciteSetBstMidEndSepPunct{\mcitedefaultmidpunct}
{\mcitedefaultendpunct}{\mcitedefaultseppunct}\relax
\EndOfBibitem
\bibitem[Zhang \latin{et~al.}(2017)Zhang, Song, Luo, Deng, Mosallanejad,
  Taniguchi, Watanabe, Li, Cao, Guo, Nori, and Guo]{zhang_electrotunable_2017}
Zhang,~Z.-Z.; Song,~X.-X.; Luo,~G.; Deng,~G.-W.; Mosallanejad,~V.;
  Taniguchi,~T.; Watanabe,~K.; Li,~H.-O.; Cao,~G.; Guo,~G.-C.; Nori,~F.;
  Guo,~G.-P. Electrotunable artificial molecules based on van der {Waals}
  heterostructures. \emph{Science Advances} \textbf{2017}, \emph{3},
  e1701699\relax
\mciteBstWouldAddEndPuncttrue
\mciteSetBstMidEndSepPunct{\mcitedefaultmidpunct}
{\mcitedefaultendpunct}{\mcitedefaultseppunct}\relax
\EndOfBibitem
\bibitem[Boddison-Chouinard \latin{et~al.}(2021)Boddison-Chouinard, Bogan,
  Fong, Watanabe, Taniguchi, Studenikin, Sachrajda, Korkusinski, Altintas,
  Bieniek, Hawrylak, Luican-Mayer, and
  Gaudreau]{boddison-chouinard_gate-controlled_2021}
Boddison-Chouinard,~J.; Bogan,~A.; Fong,~N.; Watanabe,~K.; Taniguchi,~T.;
  Studenikin,~S.; Sachrajda,~A.; Korkusinski,~M.; Altintas,~A.; Bieniek,~M.;
  Hawrylak,~P.; Luican-Mayer,~A.; Gaudreau,~L. Gate-controlled quantum dots in
  monolayer {WSe2}. \emph{Applied Physics Letters} \textbf{2021}, \emph{119},
  133104\relax
\mciteBstWouldAddEndPuncttrue
\mciteSetBstMidEndSepPunct{\mcitedefaultmidpunct}
{\mcitedefaultendpunct}{\mcitedefaultseppunct}\relax
\EndOfBibitem
\bibitem[Davari \latin{et~al.}(2020)Davari, Stacy, Mercado, Tull, Basnet,
  Pandey, Watanabe, Taniguchi, Hu, and Churchill]{davari_gate-defined_2020}
Davari,~S.; Stacy,~J.; Mercado,~A.; Tull,~J.; Basnet,~R.; Pandey,~K.;
  Watanabe,~K.; Taniguchi,~T.; Hu,~J.; Churchill,~H. Gate-{Defined}
  {Accumulation}-{Mode} {Quantum} {Dots} in {Monolayer} and {Bilayer} {W} {Se}
  2. \emph{Physical Review Applied} \textbf{2020}, \emph{13}, 054058\relax
\mciteBstWouldAddEndPuncttrue
\mciteSetBstMidEndSepPunct{\mcitedefaultmidpunct}
{\mcitedefaultendpunct}{\mcitedefaultseppunct}\relax
\EndOfBibitem
\bibitem[Papadopoulos \latin{et~al.}(2020)Papadopoulos, Gehring, Watanabe,
  Taniguchi, Van Der~Zant, and Steele]{papadopoulos_tunneling_2020}
Papadopoulos,~N.; Gehring,~P.; Watanabe,~K.; Taniguchi,~T.; Van Der~Zant,~H.
  S.~J.; Steele,~G.~A. Tunneling spectroscopy of localized states of {WS} 2
  barriers in vertical van der {Waals} heterostructures. \emph{Physical Review
  B} \textbf{2020}, \emph{101}, 165303\relax
\mciteBstWouldAddEndPuncttrue
\mciteSetBstMidEndSepPunct{\mcitedefaultmidpunct}
{\mcitedefaultendpunct}{\mcitedefaultseppunct}\relax
\EndOfBibitem
\bibitem[Devidas \latin{et~al.}(2021)Devidas, Keren, and
  Steinberg]{devidas_spectroscopy_2021}
Devidas,~T.~R.; Keren,~I.; Steinberg,~H. Spectroscopy of {NbSe} $_{\textrm{2}}$
  {Using} {Energy}-{Tunable} {Defect}-{Embedded} {Quantum} {Dots}. \emph{Nano
  Letters} \textbf{2021}, \emph{21}, 6931--6937\relax
\mciteBstWouldAddEndPuncttrue
\mciteSetBstMidEndSepPunct{\mcitedefaultmidpunct}
{\mcitedefaultendpunct}{\mcitedefaultseppunct}\relax
\EndOfBibitem
\bibitem[Kumar \latin{et~al.}()Kumar, Kim, Tripathy, Watanabe, Taniguchi,
  Novoselov, and Kotekar-Patil]{kumar_excited_nodate}
Kumar,~P.; Kim,~H.; Tripathy,~S.; Watanabe,~K.; Taniguchi,~T.;
  Novoselov,~K.~S.; Kotekar-Patil,~D. Excited state spectroscopy and spin
  splitting in atomically thin quantum dots. \emph{Nanoscale (2023)} \relax
\mciteBstWouldAddEndPunctfalse
\mciteSetBstMidEndSepPunct{\mcitedefaultmidpunct}
{}{\mcitedefaultseppunct}\relax
\EndOfBibitem
\bibitem[Hofheinz \latin{et~al.}(2006)Hofheinz, Jehl, Sanquer, Molas, Vinet,
  and Deleonibus]{hofheinz_simple_2006}
Hofheinz,~M.; Jehl,~X.; Sanquer,~M.; Molas,~G.; Vinet,~M.; Deleonibus,~S.
  Simple and controlled single electron transistor based on doping modulation
  in silicon nanowires. \emph{Applied Physics Letters} \textbf{2006},
  \emph{89}, 143504\relax
\mciteBstWouldAddEndPuncttrue
\mciteSetBstMidEndSepPunct{\mcitedefaultmidpunct}
{\mcitedefaultendpunct}{\mcitedefaultseppunct}\relax
\EndOfBibitem
\bibitem[Leobandung \latin{et~al.}(1995)Leobandung, Guo, Wang, and
  Chou]{leobandung_observation_1995}
Leobandung,~E.; Guo,~L.; Wang,~Y.; Chou,~S.~Y. Observation of quantum effects
  and {Coulomb} blockade in silicon quantum-dot transistors at temperatures
  over 100 {K}. \emph{Applied Physics Letters} \textbf{1995}, \emph{67},
  938--940\relax
\mciteBstWouldAddEndPuncttrue
\mciteSetBstMidEndSepPunct{\mcitedefaultmidpunct}
{\mcitedefaultendpunct}{\mcitedefaultseppunct}\relax
\EndOfBibitem
\bibitem[Angus \latin{et~al.}(2007)Angus, Ferguson, Dzurak, and
  Clark]{angus_gate-defined_2007}
Angus,~S.~J.; Ferguson,~A.~J.; Dzurak,~A.~S.; Clark,~R.~G. Gate-{Defined}
  {Quantum} {Dots} in {Intrinsic} {Silicon}. \emph{Nano Letters} \textbf{2007},
  \emph{7}, 2051--2055\relax
\mciteBstWouldAddEndPuncttrue
\mciteSetBstMidEndSepPunct{\mcitedefaultmidpunct}
{\mcitedefaultendpunct}{\mcitedefaultseppunct}\relax
\EndOfBibitem
\bibitem[Korm\'anyos \latin{et~al.}(2014)Korm\'anyos, Z\'olyomi, Drummond, and
  Burkard]{Burkard2014}
Korm\'anyos,~A.; Z\'olyomi,~V.; Drummond,~N.~D.; Burkard,~G. Spin-Orbit
  Coupling, Quantum Dots, and Qubits in Monolayer Transition Metal
  Dichalcogenides. \emph{Phys. Rev. X} \textbf{2014}, \emph{4}, 011034\relax
\mciteBstWouldAddEndPuncttrue
\mciteSetBstMidEndSepPunct{\mcitedefaultmidpunct}
{\mcitedefaultendpunct}{\mcitedefaultseppunct}\relax
\EndOfBibitem
\bibitem[Liu \latin{et~al.}(2014)Liu, Pang, Yao, and Yao]{Liu2014}
Liu,~G.-B.; Pang,~H.; Yao,~Y.; Yao,~W. Intervalley coupling by quantum dot
  confinement potentials in monolayer transition metal dichalcogenides.
  \emph{New Journal of Physics} \textbf{2014}, \emph{16}, 105011\relax
\mciteBstWouldAddEndPuncttrue
\mciteSetBstMidEndSepPunct{\mcitedefaultmidpunct}
{\mcitedefaultendpunct}{\mcitedefaultseppunct}\relax
\EndOfBibitem
\bibitem[Ko\ifmmode~\acute{s}\else \'{s}\fi{}mider
  \latin{et~al.}(2013)Ko\ifmmode~\acute{s}\else \'{s}\fi{}mider, Gonz\'alez,
  and Fern\'andez-Rossier]{Kosmider2013}
Ko\ifmmode~\acute{s}\else \'{s}\fi{}mider,~K.; Gonz\'alez,~J.~W.;
  Fern\'andez-Rossier,~J. Large spin splitting in the conduction band of
  transition metal dichalcogenide monolayers. \emph{Phys. Rev. B}
  \textbf{2013}, \emph{88}, 245436\relax
\mciteBstWouldAddEndPuncttrue
\mciteSetBstMidEndSepPunct{\mcitedefaultmidpunct}
{\mcitedefaultendpunct}{\mcitedefaultseppunct}\relax
\EndOfBibitem
\bibitem[Paw\l{}owski \latin{et~al.}(2016)Paw\l{}owski, Szumniak, and
  Bednarek]{Pawlowski2016}
Paw\l{}owski,~J.; Szumniak,~P.; Bednarek,~S. Electron spin rotations induced by
  oscillating Rashba interaction in a quantum wire. \emph{Phys. Rev. B}
  \textbf{2016}, \emph{93}, 045309\relax
\mciteBstWouldAddEndPuncttrue
\mciteSetBstMidEndSepPunct{\mcitedefaultmidpunct}
{\mcitedefaultendpunct}{\mcitedefaultseppunct}\relax
\EndOfBibitem
\bibitem[Laturia \latin{et~al.}(2018)Laturia, Van~de Put, and
  Vandenberghe]{Laturia2018}
Laturia,~A.; Van~de Put,~M.~L.; Vandenberghe,~W.~G. Dielectric properties of
  hexagonal boron nitride and transition metal dichalcogenides: from monolayer
  to bulk. \emph{npj 2D Materials and Applications} \textbf{2018}, \emph{2},
  6\relax
\mciteBstWouldAddEndPuncttrue
\mciteSetBstMidEndSepPunct{\mcitedefaultmidpunct}
{\mcitedefaultendpunct}{\mcitedefaultseppunct}\relax
\EndOfBibitem
\bibitem[Kouwenhoven \latin{et~al.}(2001)Kouwenhoven, Austing, and
  Tarucha]{Kouwenhoven_2001}
Kouwenhoven,~L.~P.; Austing,~D.~G.; Tarucha,~S. Few-electron quantum dots.
  \emph{Reports on Progress in Physics} \textbf{2001}, \emph{64}, 701\relax
\mciteBstWouldAddEndPuncttrue
\mciteSetBstMidEndSepPunct{\mcitedefaultmidpunct}
{\mcitedefaultendpunct}{\mcitedefaultseppunct}\relax
\EndOfBibitem
\bibitem[David \latin{et~al.}(2018)David, Burkard, and
  Kormányos]{david_effective_2018}
David,~A.; Burkard,~G.; Kormányos,~A. Effective theory of monolayer {TMDC}
  double quantum dots. \emph{2D Materials} \textbf{2018}, \emph{5},
  035031\relax
\mciteBstWouldAddEndPuncttrue
\mciteSetBstMidEndSepPunct{\mcitedefaultmidpunct}
{\mcitedefaultendpunct}{\mcitedefaultseppunct}\relax
\EndOfBibitem
\bibitem[Freitag \latin{et~al.}(2016)Freitag, Chizhova, Nemes-Incze, Woods,
  Gorbachev, Cao, Geim, Novoselov, Burgd{\"o}rfer, Libisch, and
  Morgenstern]{freitag}
Freitag,~N.~M.; Chizhova,~L.~A.; Nemes-Incze,~P.; Woods,~C.~R.;
  Gorbachev,~R.~V.; Cao,~Y.; Geim,~A.~K.; Novoselov,~K.~S.; Burgd{\"o}rfer,~J.;
  Libisch,~F.; Morgenstern,~M. Electrostatically Confined Monolayer Graphene
  Quantum Dots with Orbital and Valley Splittings. \emph{Nano Letters}
  \textbf{2016}, \emph{16}, 5798--5805\relax
\mciteBstWouldAddEndPuncttrue
\mciteSetBstMidEndSepPunct{\mcitedefaultmidpunct}
{\mcitedefaultendpunct}{\mcitedefaultseppunct}\relax
\EndOfBibitem
\bibitem[Kretinin \latin{et~al.}(2010)Kretinin, Popovitz-Biro, Mahalu, and
  Shtrikman]{kretinin}
Kretinin,~A.~V.; Popovitz-Biro,~R.; Mahalu,~D.; Shtrikman,~H. Multimode
  Fabry-P{\'e}rot Conductance Oscillations in Suspended Stacking-Faults-Free
  InAs Nanowires. \emph{Nano Letters} \textbf{2010}, \emph{10},
  3439--3445\relax
\mciteBstWouldAddEndPuncttrue
\mciteSetBstMidEndSepPunct{\mcitedefaultmidpunct}
{\mcitedefaultendpunct}{\mcitedefaultseppunct}\relax
\EndOfBibitem
\bibitem[Cui \latin{et~al.}(2015)Cui, Lee, Kim, Arefe, Huang, Lee, Chenet,
  Zhang, Wang, Ye, Pizzocchero, Jessen, Watanabe, Taniguchi, Muller, Low, Kim,
  and Hone]{cui_multi-terminal_2015}
Cui,~X. \latin{et~al.}  Multi-terminal transport measurements of {MoS2} using a
  van der {Waals} heterostructure device platform. \emph{Nature Nanotechnology}
  \textbf{2015}, \emph{10}, 534--540\relax
\mciteBstWouldAddEndPuncttrue
\mciteSetBstMidEndSepPunct{\mcitedefaultmidpunct}
{\mcitedefaultendpunct}{\mcitedefaultseppunct}\relax
\EndOfBibitem
\bibitem[Cui \latin{et~al.}(2017)Cui, Shih, Jauregui, Chae, Kim, Li, Seo,
  Pistunova, Yin, Park, Choi, Lee, Watanabe, Taniguchi, Kim, Dean, and
  Hone]{cui_low-temperature_2017}
Cui,~X. \latin{et~al.}  Low-{Temperature} {Ohmic} {Contact} to {Monolayer}
  {MoS} $_{\textrm{2}}$ by van der {Waals} {Bonded} {Co}/ \textit{h} -{BN}
  {Electrodes}. \emph{Nano Letters} \textbf{2017}, \emph{17}, 4781--4786\relax
\mciteBstWouldAddEndPuncttrue
\mciteSetBstMidEndSepPunct{\mcitedefaultmidpunct}
{\mcitedefaultendpunct}{\mcitedefaultseppunct}\relax
\EndOfBibitem
\bibitem[Liu \latin{et~al.}(2018)Liu, Guo, Zhu, Liao, Lee, Ding, Shakir,
  Gambin, Huang, and Duan]{liu_approaching_2018}
Liu,~Y.; Guo,~J.; Zhu,~E.; Liao,~L.; Lee,~S.-J.; Ding,~M.; Shakir,~I.;
  Gambin,~V.; Huang,~Y.; Duan,~X. Approaching the {Schottky}–{Mott} limit in
  van der {Waals} metal–semiconductor junctions. \emph{Nature} \textbf{2018},
  \emph{557}, 696--700\relax
\mciteBstWouldAddEndPuncttrue
\mciteSetBstMidEndSepPunct{\mcitedefaultmidpunct}
{\mcitedefaultendpunct}{\mcitedefaultseppunct}\relax
\EndOfBibitem
\bibitem[Wang \latin{et~al.}(2019)Wang, Kim, Wu, Martinez, Song, Yang, Zhao,
  Mkhoyan, Jeong, and Chhowalla]{wang_van_2019}
Wang,~Y.; Kim,~J.~C.; Wu,~R.~J.; Martinez,~J.; Song,~X.; Yang,~J.; Zhao,~F.;
  Mkhoyan,~A.; Jeong,~H.~Y.; Chhowalla,~M. Van der {Waals} contacts between
  three-dimensional metals and two-dimensional semiconductors. \emph{Nature}
  \textbf{2019}, \emph{568}, 70--74\relax
\mciteBstWouldAddEndPuncttrue
\mciteSetBstMidEndSepPunct{\mcitedefaultmidpunct}
{\mcitedefaultendpunct}{\mcitedefaultseppunct}\relax
\EndOfBibitem
\bibitem[Shen \latin{et~al.}(2021)Shen, Su, Lin, Chou, Cheng, Park, Chiu, Lu,
  Tang, Tavakoli, Pitner, Ji, Cai, Mao, Wang, Tung, Li, Bokor, Zettl, Wu,
  Palacios, Li, and Kong]{shen_ultralow_2021}
Shen,~P.-C. \latin{et~al.}  Ultralow contact resistance between semimetal and
  monolayer semiconductors. \emph{Nature} \textbf{2021}, \emph{593},
  211--217\relax
\mciteBstWouldAddEndPuncttrue
\mciteSetBstMidEndSepPunct{\mcitedefaultmidpunct}
{\mcitedefaultendpunct}{\mcitedefaultseppunct}\relax
\EndOfBibitem
\bibitem[Pisoni \latin{et~al.}(2018)Pisoni, Kormányos, Brooks, Lei, Back,
  Eich, Overweg, Lee, Rickhaus, Watanabe, Taniguchi, Imamoglu, Burkard, Ihn,
  and Ensslin]{pisoni_interactions_2018}
Pisoni,~R.; Kormányos,~A.; Brooks,~M.; Lei,~Z.; Back,~P.; Eich,~M.;
  Overweg,~H.; Lee,~Y.; Rickhaus,~P.; Watanabe,~K.; Taniguchi,~T.;
  Imamoglu,~A.; Burkard,~G.; Ihn,~T.; Ensslin,~K. Interactions and
  {Magnetotransport} through {Spin}-{Valley} {Coupled} {Landau} {Levels} in
  {Monolayer} {MoS} 2. \emph{Physical Review Letters} \textbf{2018},
  \emph{121}, 247701\relax
\mciteBstWouldAddEndPuncttrue
\mciteSetBstMidEndSepPunct{\mcitedefaultmidpunct}
{\mcitedefaultendpunct}{\mcitedefaultseppunct}\relax
\EndOfBibitem
\bibitem[Wang \latin{et~al.}(2013)Wang, Meric, Huang, Gao, Gao, Tran,
  Taniguchi, Watanabe, Campos, Muller, Guo, Kim, Hone, Shepard, and
  Dean]{wang_one-dimensional_2013}
Wang,~L.; Meric,~I.; Huang,~P.~Y.; Gao,~Q.; Gao,~Y.; Tran,~H.; Taniguchi,~T.;
  Watanabe,~K.; Campos,~L.~M.; Muller,~D.~A.; Guo,~J.; Kim,~P.; Hone,~J.;
  Shepard,~K.~L.; Dean,~C.~R. One-{Dimensional} {Electrical} {Contact} to a
  {Two}-{Dimensional} {Material}. \emph{Science} \textbf{2013}, \emph{342},
  614--617\relax
\mciteBstWouldAddEndPuncttrue
\mciteSetBstMidEndSepPunct{\mcitedefaultmidpunct}
{\mcitedefaultendpunct}{\mcitedefaultseppunct}\relax
\EndOfBibitem
\bibitem[Cohen \latin{et~al.}(2023)Cohen, Samuelson, Wang, Klocke, Reeves,
  Taniguchi, Watanabe, Vijay, Zaletel, and Young]{cohen_nanoscale_2023}
Cohen,~L.~A.; Samuelson,~N.~L.; Wang,~T.; Klocke,~K.; Reeves,~C.~C.;
  Taniguchi,~T.; Watanabe,~K.; Vijay,~S.; Zaletel,~M.~P.; Young,~A.~F.
  Nanoscale electrostatic control in ultraclean van der {Waals}
  heterostructures by local anodic oxidation of graphite gates. \emph{Nat.
  Phys.} \textbf{2023}, \emph{19}, 1502--1508\relax
\mciteBstWouldAddEndPuncttrue
\mciteSetBstMidEndSepPunct{\mcitedefaultmidpunct}
{\mcitedefaultendpunct}{\mcitedefaultseppunct}\relax
\EndOfBibitem
\end{mcitethebibliography}

\end{document}


\section{Other device setups}
Below we present Poisson-Schr\"odindger calculations for some other configurations and compare them showing why in the final proposition we select the Configuration~C -- one presented in the main text. In all configurations we assume the thickness of the tunnel barriers (TBs) as $20$~nm.

Configuration~A, with the fine gates (FG) and the split gates (SG) above the TMDC layer. Voltages are applied to get a single QD confinement below the central fine gate: $V_L=-100$, $V_C=-500$, $V_R=-100$, $V_\mathrm{SG}=100$, $V_\mathrm{back}=-400$~(mV). When solving the Poisson equation the mesh grid should be carefully selected to properly describe potential details next to these (fine, spilt, and monolayer) layers. For such a configuration the confinement is a bit asymmetric. Higher SG-gap (e.g. $100$~nm) will make confinement almost symmetric. Possible issues with this device setup: fine gates are screened by the split gates thus it might be hard to get double QD structure in this geometry. Moreover, QD separation from the electron gas might be too weak and a barrier that controls coupling to the gas hard to control. The schematic of the device in the configuration~A and the results for the Poisson-Schr\"odindger equation are shown in Figure~S1.

Configuration~B with higher voltages at $V_L$, $V_R$, and split gates to repel the electron more strongly (and lower backgate voltage to equalize this). Assumed voltages are: $V_L=300$, $V_C=-300$, $V_R=300$, $V_\mathrm{SG}=500$, $V_\mathrm{back}=-2000$~(mV). We observe comparable splittings as for Configuration~A, but very different behavior of splitting vs. SG gap -- here a lower SG gaps are optimal (give higher splitting) than for Configuration~A. Further increasing of the repulsing voltages does not increase the confinement. The schematic of the device in the configuration~B and the results for the Poisson-Schr\"odindger equation are shown in Figure~S2.

Configuration~C, one analyzed in the main text: voltages same as for Configuration~B. This is the best configuration, with the highest splittings.
The comparison of splittings for the various configurations is presented in Figure~S5.
The schematic of the device in the configuration~C and the results for the Poisson-Schr\"odindger equation are shown in Figure~S3.

Configuration~C1, same as Configuration~C but with double QD structure created. Now the central fine gate is repelling, and $V_L$, $V_R$ attracting: $V_L=-250$, $V_C=200$, $V_R=-350$,$V_\mathrm{split}=500$, $V_\mathrm{back}=-2000$~(mV). When we strongly decouple the both dots splittings are even larger. For weaker decoupling (lower $V_C$), the splittings are similar like in Configuration~C. The schematic of the device in the configuration~C1 and the results for the Poisson-Schr\"odindger equation are shown in Figure~S4.

\begin{figure}
\includegraphics[width=10cm]{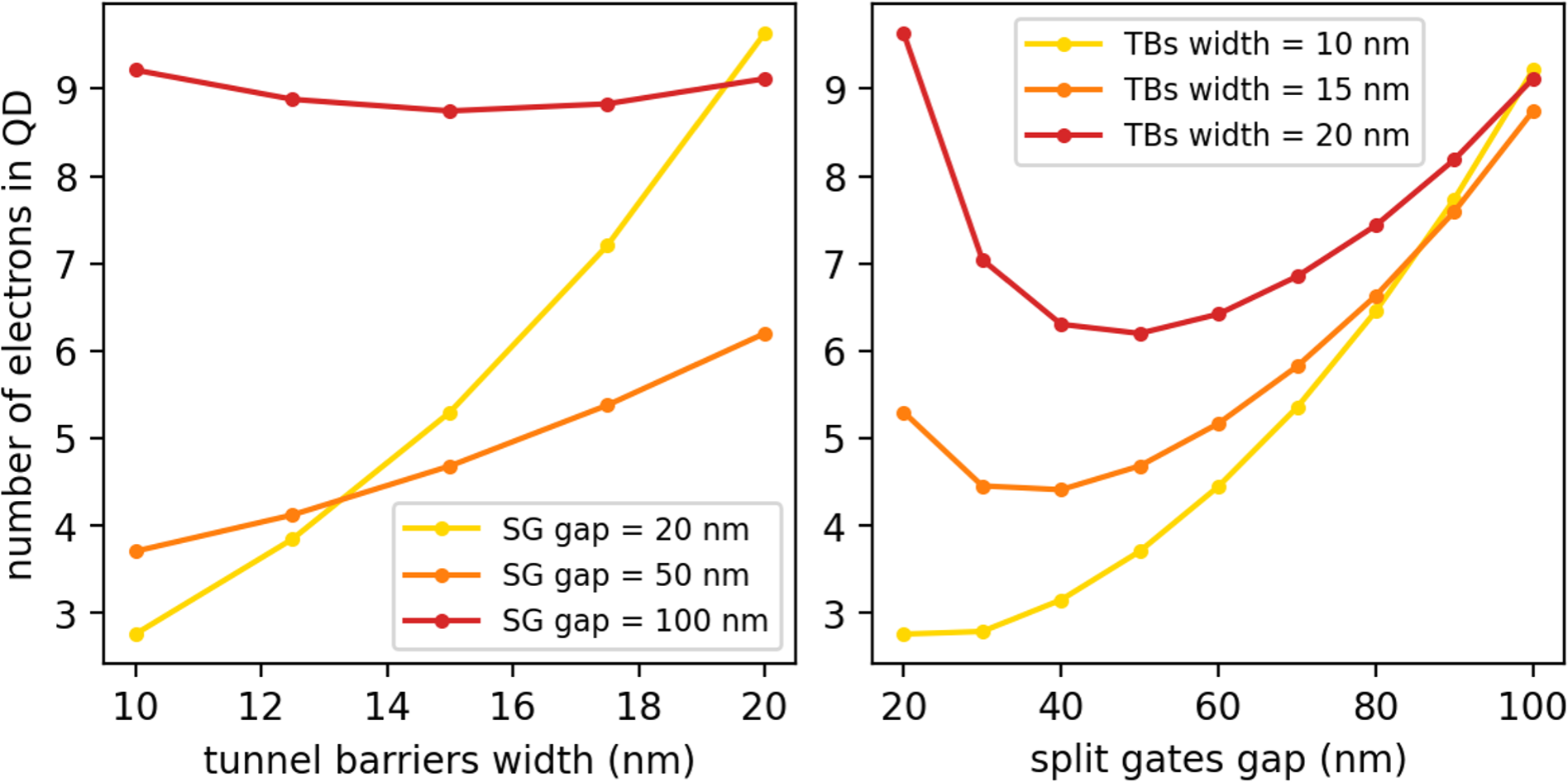}
\caption{\label{fig:s6} Estimation of the number of electrons confined within the quantum dot for different configurations of the tunnel barriers width and split gates gap.
}
\centering
\end{figure}

Additionally, in Figure~\ref{fig:s6} we present results for the estimation of the number of electrons confined within the QD (for different nanodevice geometries, controlled by the TBs width and SG gap) at the Fermi level and electron gas density resulting from the self-consistent Poisson-Schr\"odindger calculations. In the calculations a non-interacting Fermi gas model was assumed.

\begin{figure}[tb]
\includegraphics[width=15cm]{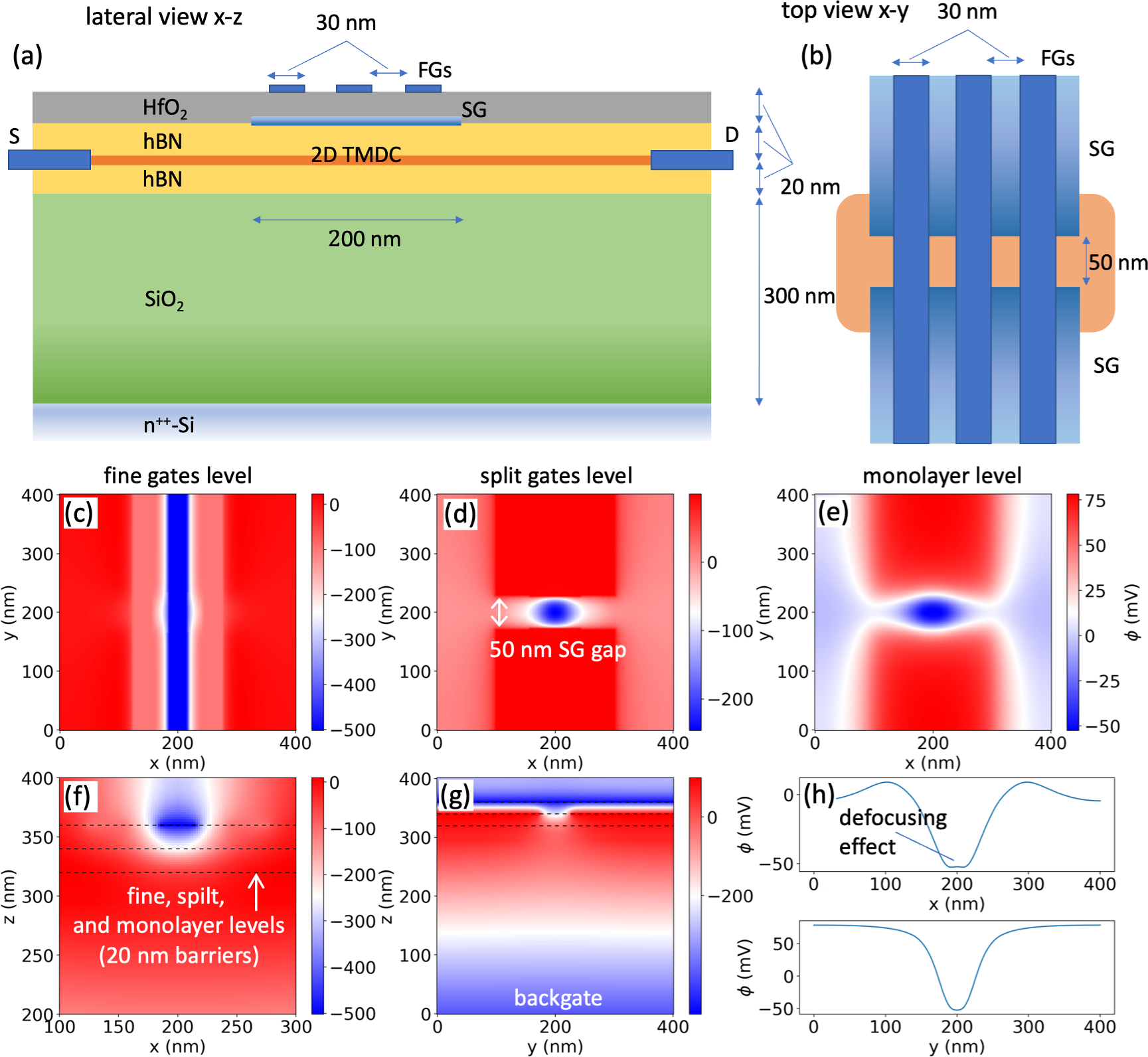}
\caption{\label{fig:s1} Detailed results for Configuration~A: scheme showing the device structure from (a) lateral and (b) top view; (c-h) detailed electrostatic potential profiles from the Poisson-Schr\"odinger calculations for different cross-cuts.}
\centering
\end{figure}

\begin{figure}
\includegraphics[width=15cm]{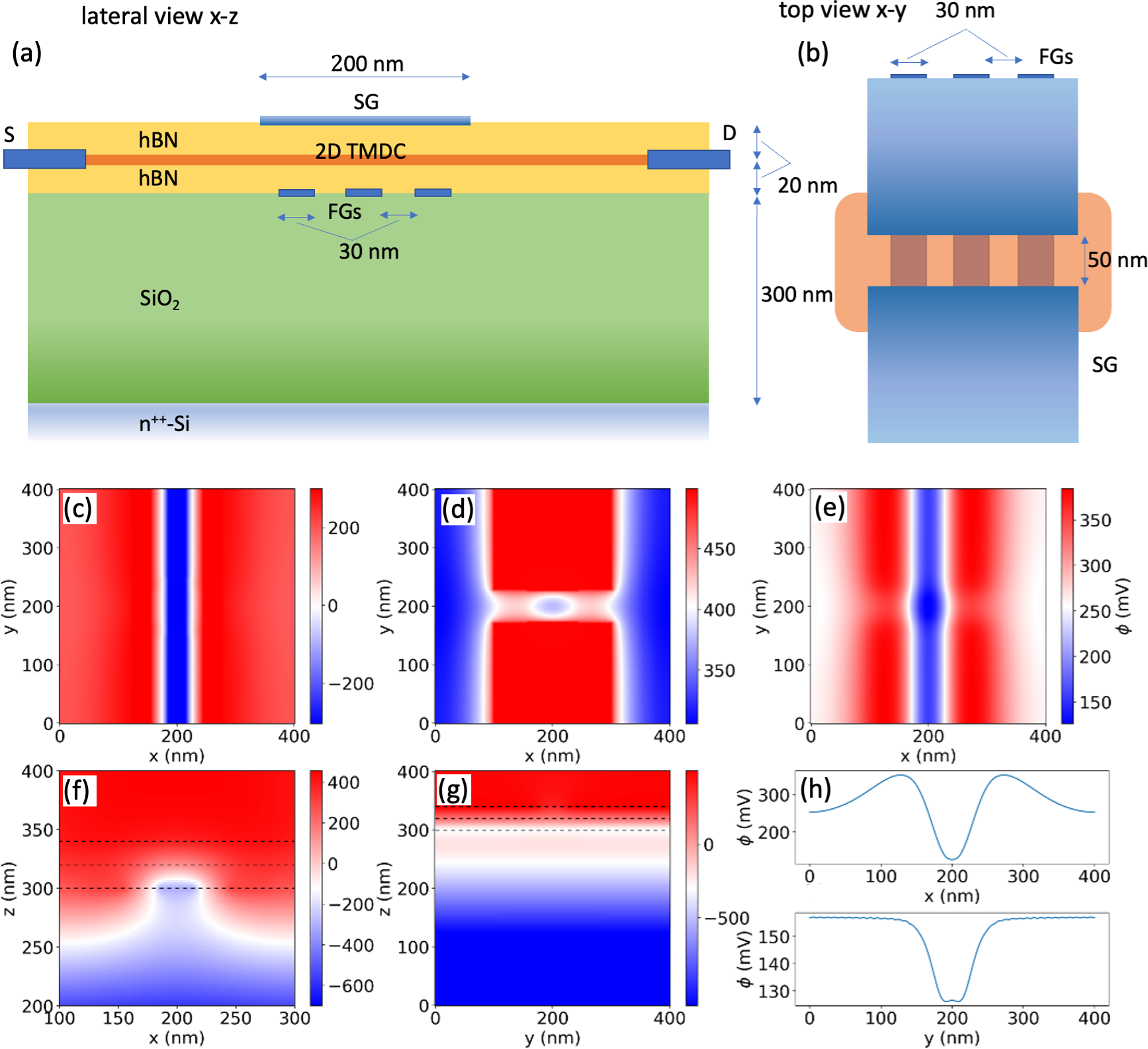}
\caption{\label{fig:s2} Configuration B. Caption similar as for Figure~S1.}
\centering
\end{figure}

\begin{figure}
\includegraphics[width=15cm]{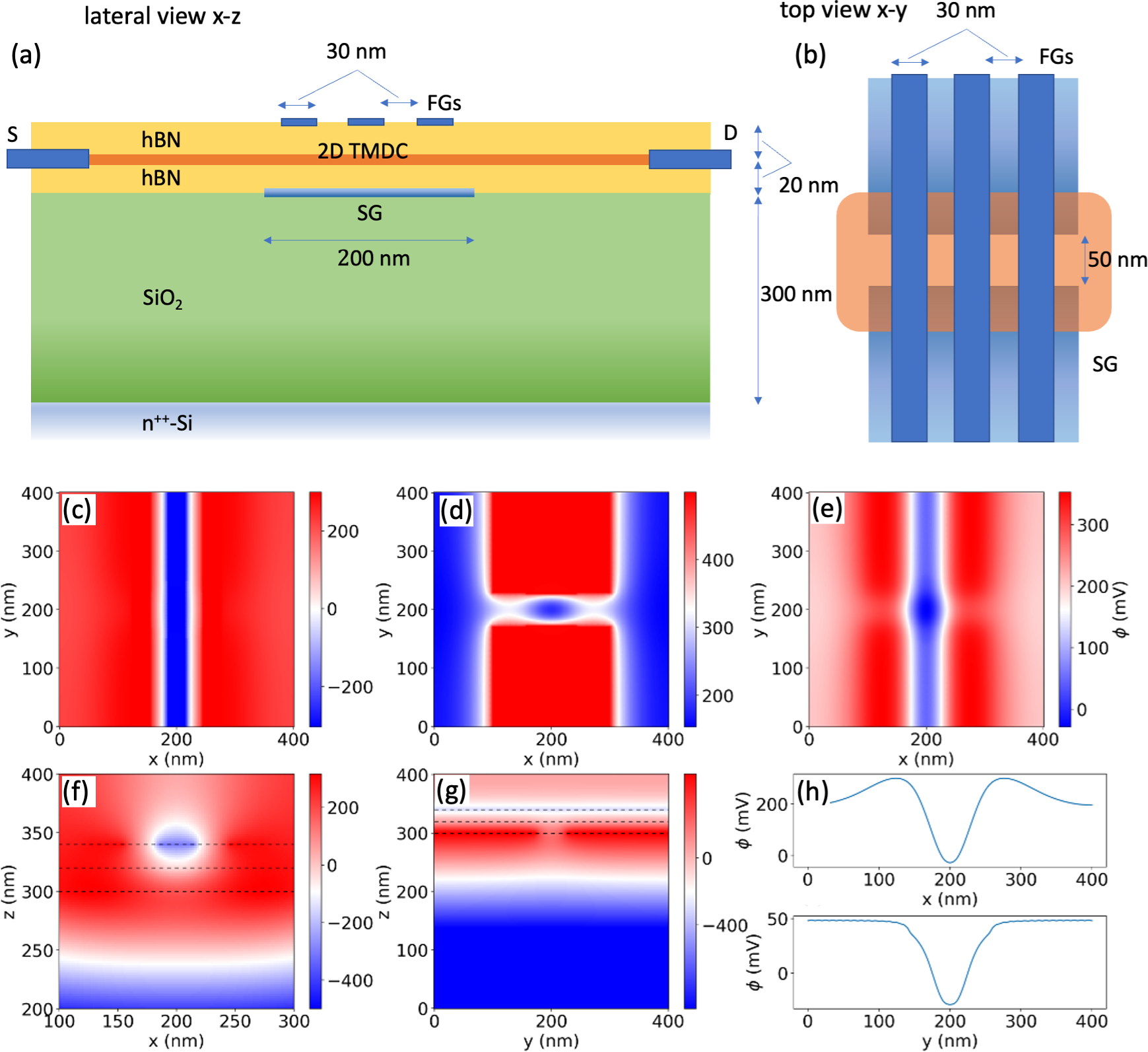}
\caption{\label{fig:s3} Configuration C. Caption similar as for Figure~S1.}
\centering
\end{figure}

\begin{figure}
\includegraphics[width=15cm]{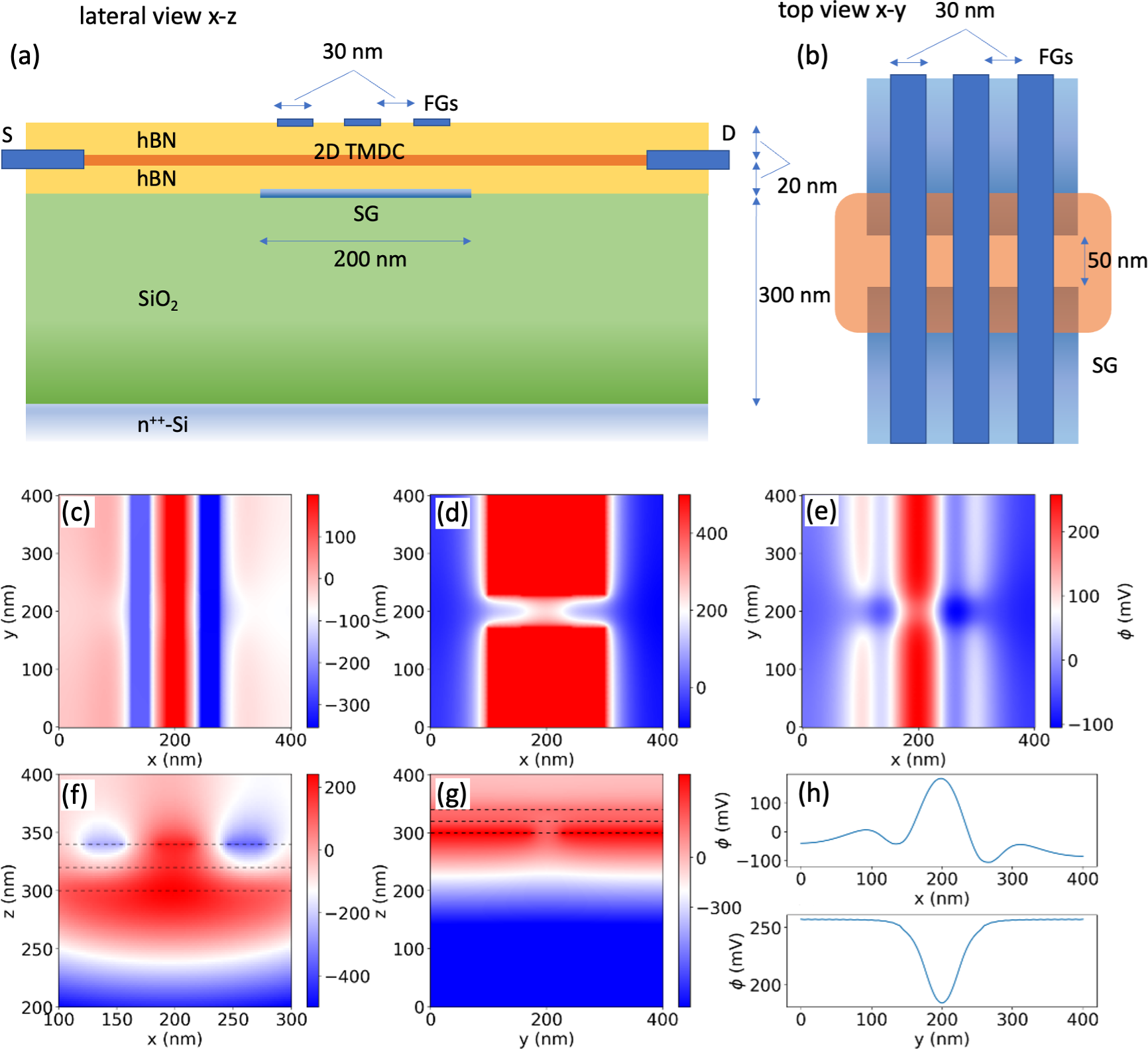}
\caption{\label{fig:s4} Configuration C1. Caption similar as for Figure~S1.}
\centering
\end{figure}

\begin{figure}
\includegraphics[width=16cm]{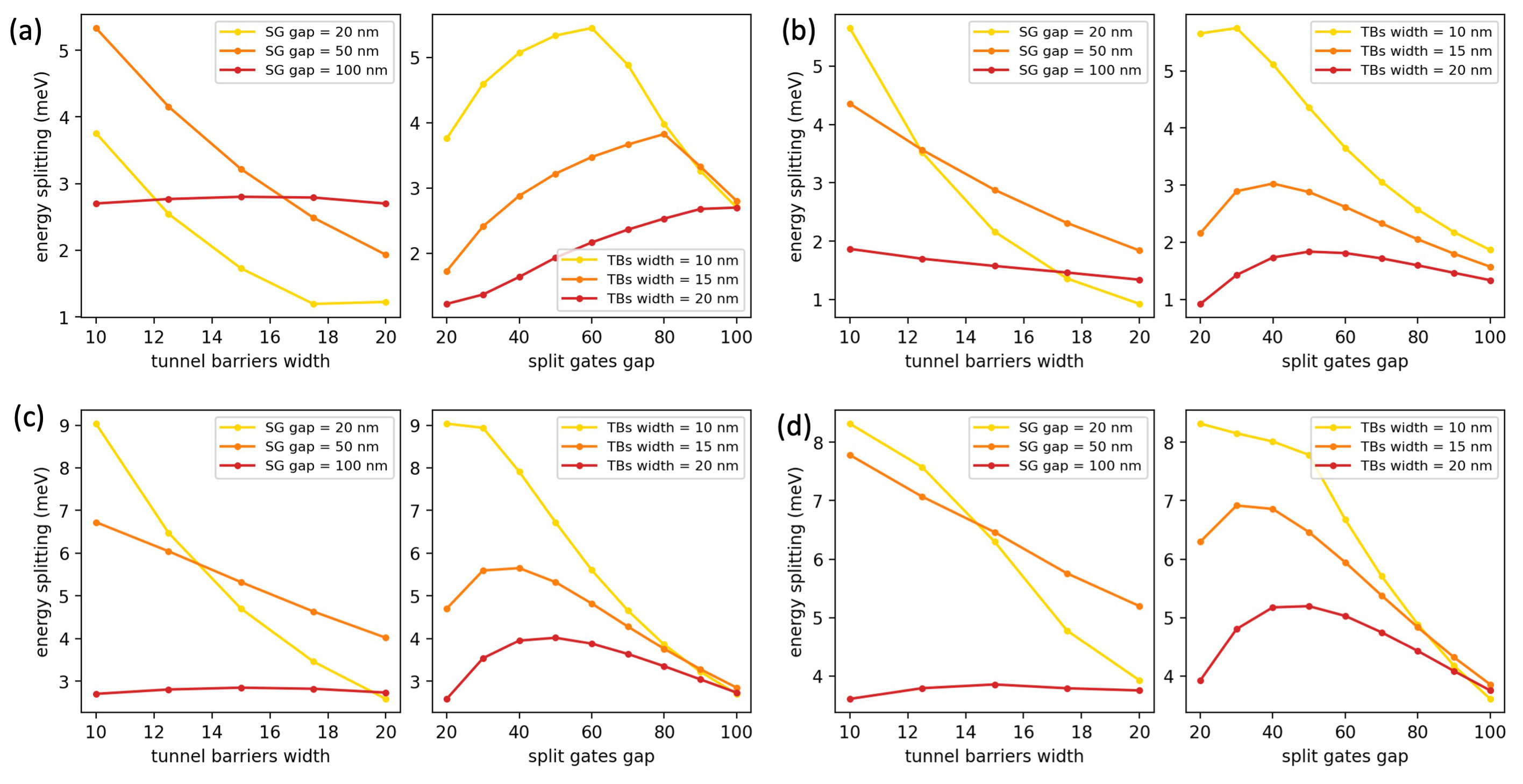}
\caption{\label{fig:s5} Comparison of splittings for various configurations. (a) Configuration~A: Splitting @$(50,20)=1.93$~meV (notation $(50,20)$ means SG gap $=50$~nm and TBs width $=20$~nm). Starting from a certain value of the SG gap (about $100$~nm), tunnel barriers width became insignificant on splitting. (b) Configuration~B: splitting @$(50,20)=1.84$~meV. Also here, for wide SG gap, tunnel barriers width became insignificant on splitting. (c) Configuration~C, same as in the main text, with much higher splitting @$(50,20)=4.02$~meV. Configuration~C1, same as C but with a double dot structure, now splitting @$(50,20)=5.19$~meV for the lower dot.
}
\centering
\end{figure}

\begin{figure}
\includegraphics[width=15cm]{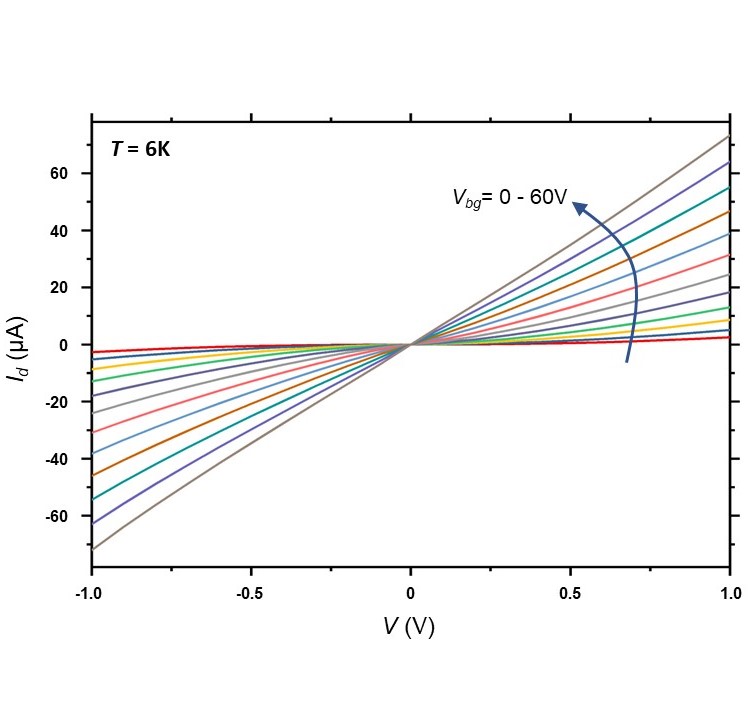}
\caption{\label{fig:s7} Output transfer characteristic in a device with single layer MoS$_2$ contacted with Bismuth. Different color traces are taken at varying $V_\mathrm{bg}$ from $0$~V to $60$~V and $T=6$~K. Linear $I_d$--$V$ is an indication of Ohmic contact.}
\centering
\end{figure}